\documentclass[aps,floatfix,superscriptaddress]{revtex4}
\usepackage{amsmath,amssymb,eucal,graphicx,subfigure}
\usepackage{color}
\usepackage{mathtools,bm,soul}
\begin{document}

\newcommand{\cov}{\operatornamewithlimits{Cov}}

\title{Identifying Coupling Structure in Complex Systems through the Optimal Causation Entropy Principle}

\author{Jie Sun}
\email{sunj@clarkson.edu}
\affiliation{Department of Mathematics, Clarkson University, Potsdam, NY 13699-5815, USA}
\author{Carlo Cafaro}
\email{ccafaro@clarkson.edu}
\affiliation{Department of Mathematics, Clarkson University, Potsdam, NY 13699-5815, USA}
\author{Erik M. Bollt}
\email{bolltem@clarkson.edu}
\affiliation{Department of Mathematics, Clarkson University, Potsdam, NY 13699-5815, USA}

\begin{abstract}
Inferring the coupling structure of complex systems from time series data in general by means of statistical and information-theoretic techniques is a challenging problem in applied science. The reliability of statistical inferences requires the construction of suitable information-theoretic measures that take into account both direct and indirect influences, manifest in the form of information flows, between the components within the system. In this work, we present an application of the optimal causation entropy (oCSE) principle to identify the coupling structure of a synthetic biological system, the repressilator. Specifically, when the system reaches an equilibrium state, we use a stochastic perturbation approach to extract time series data that approximate a linear stochastic process. Then, we present and jointly apply the aggregative discovery and progressive removal algorithms based on the oCSE principle to infer the coupling structure of the system from the measured data. Finally, we show that the success rate of our coupling inferences not only improves with the amount of available data, but it also increases with a higher frequency of sampling and is especially immune to false positives.
\end{abstract}

\maketitle

\section{Introduction}
Deducing equations of dynamics from empirical observations is fundamental in science. In real-world experiments, we gather data of the state of a system. Then, to achieve the comprehension of the mechanisms behind the system dynamics, we often need to reconstruct the underlying dynamical equations from the measured data. For example, the laws of celestial mechanics were deduced based on observations of planet trajectories~\cite{Fitzpatrick}; the forms of chemical equations were inferred upon empirical reaction relations and kinetics~\cite{Atkins}; the principles of economics were uncovered through market data analysis~\cite{Mankiw}.
Despite such important accomplishments, the general problem of identifying dynamical equations from data is a challenging one.
Early efforts, such as the one presented in \cite{Crutchfield1987}, utilized embedding theory and relative entropy to reconstruct the deterministic component of a low-dimensional dynamical system.

Later, more systematic methods were developed to design optimal models from a set of basis functions, where model quality was quantified in various ways, such as the Euclidean norm of the error~\cite{Yao2007}, the length of the prediction time window (chaotic shadowing)~\cite{Sun2011} or the sparsity in the model terms (compressive sensing)~\cite{Wang2011}. Each method achieved success in a range of systems, but none is generally applicable. This is, in fact, not surprising in light of the recent work that showed that the identification of exact dynamical equations from data is NP hard and, therefore, unlikely to be solved efficiently~\cite{Cubitt2012}.

Fortunately, in many applications, the problem we face is not necessarily the extraction of exact equations, but rather, uncovering the cause-and-effect relationships (\emph{i.e.}, direct coupling structure) among the components within a complex system. For example, in medical diagnosis, the primary goal is to identify the causes of a disease and/or the roots of a disorder, so as to prescribe effective treatments. In structural health monitoring, the main objective is to locate the defects that could cause abrupt changes of the connectivity structure and adverse the performance of the system. Due to its wide range of applications, the problem of inferring causal relationships from observational data has attracted broad attention over the past few decades~\cite{Heider1944,Granger1969,Granger1988,Spirtes2000,Rothman2005,Caticha2007,Frenzel2007,Schindlera2007,Guo2008,Heckman2008,Pearl2009,Gao2011,Vicente2011,Runge2012PRL,Runge2012PRE,Sun2014PhysicaD,Sun2014arXiv}.

Among the various notions of causality, we adopt the one originally proposed by Granger, which relies on two basic principles~\cite{Granger1969,Granger1988}:\vspace{6pt}
\begin{enumerate}\vspace{-0.1in}
\item[(1)] The cause should occur before the effect (caused);\vspace{-0.1in}\vspace{-3pt}
\item[(2)] The causal process should carry information (unavailable in other processes) about the effect.
\end{enumerate}\vspace{-0.1in}
\vspace{6pt}

\begin{figure}[htbp]
\centering
\includegraphics*[width=0.8\textwidth]{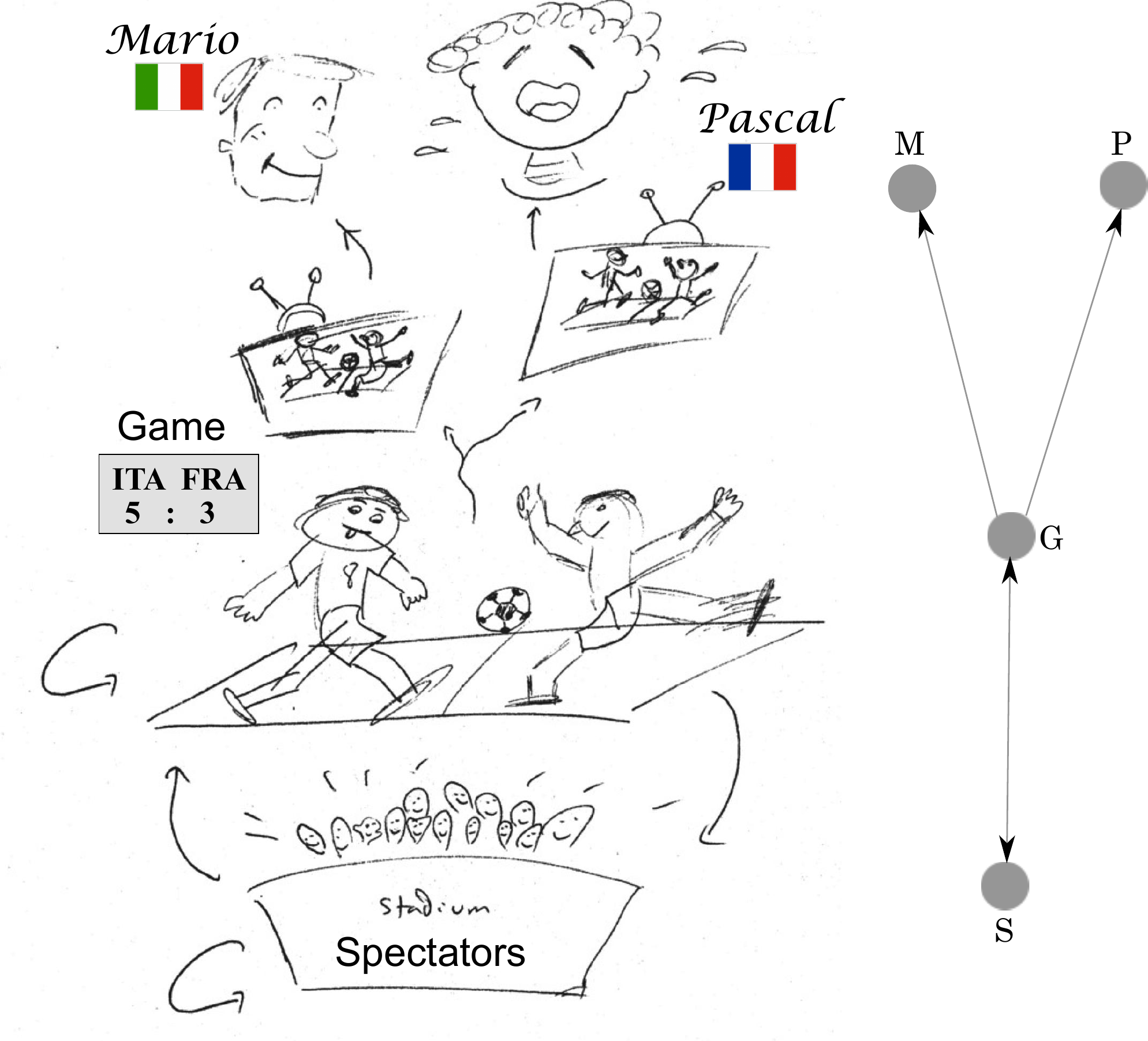}
\caption{\baselineskip 14pt
Cause-and-effect relationships during a soccer game.\\
The story: It is July 9, 2006. Nearly $750$ million people are watching the FIFA World Cup final between Italy and France.
The soccer game is held at the Olympiastadion in Berlin in the presence of around $69,000$ live spectators.
Mario is watching the game in a hotel, cheering for Italy. Pascal is sitting comfortably in front of his new TV, supporting France.\\
The Causality Quiz: answer ``Yes'' or ``No'' to the following questions, and explain why.\\
1. What affects the state of mind of Mario?\\
$\Box$~Is Mario happy because Pascal is sad? {\scriptsize No. Mario has no idea about who Pascal is.}\\
$\Box$~Is Mario happy because the spectators are cheering? {\scriptsize No. If anything, Mario is only jealous of those attending the game.}\\
$\Box$~Is Mario happy because of the game? {\scriptsize Yes. Check the scoreboard.}\\
2. What affects the behavior of the spectators?\\
$\Box$~Are the spectators cheering because Mario is happy? {\scriptsize No. Why would they?}\\
$\Box$~Are the spectators cheering because Pascal is sad? {\scriptsize No. Why would they?}\\
$\Box$~Are the spectators cheering because of the game? {\scriptsize Yes. They are restless soccer lovers, just like the players.}\\
3. What affects the state of the game?\\
$\Box$~Is Mario helping his team to win? {\scriptsize No. Although Mario probably thinks so after too much wine and cheese.}\\
$\Box$~Is Pascal causing his team to lose? {\scriptsize No. Pascal is only causing his TV to break after kicking a ball against it.}\\
$\Box$~Do the spectators influence the game? {\scriptsize Yes. This is even scientifically proven~\cite{Nevil2002}.}\\
The actual coupling structure: The four components in the system are Mario (M), Pascal (P), the game (G) and the spectators (S). They form a directed network of four nodes, as shown on the right of the picture (with self-links ignored). The game affects the state of mind of Mario, Pascal and the spectators. On the other hand, the spectators influence the game.
}\label{fig1}
\end{figure}

A causal relationship needs to satisfy both requirements. See Figure~\ref{fig1} as a schematic illustration.

The classical Granger causality is limited to coupled linear systems, while most recently developed methods based on information-theoretic measures are applicable to virtually any model, although their effectiveness relies on the abundance of data. Notably, transfer entropy (TE), a particular type of conditional mutual information, was introduced to quantify the (asymmetric) information flow between pairs of components within a system~\cite{Schreiber2000,Kaiser2002} and further developed to detect the directionality of coupling in systems with two components~\cite{Palus2001,Vejmelka2008,Bollt2012}.
The application of these pairwise inference methods to identify direct coupling in systems with more than two components warrants caution: the inferred couplings are often incorrect regardless of the amount of available data~\cite{Smirnov2013,Sun2014PhysicaD,Sun2014arXiv}.
In fact, Granger's second principle implies that a true causal relationship should remain effective upon the removal of all other possible causes.
As a consequence, an inference method cannot correctly identify the direct coupling between two components in a system without appropriate conditioning on the remaining components~\cite{Frenzel2007,Sun2014PhysicaD,Sun2014arXiv}.

Inferring causal relationships in large-scale complex systems is challenging. For a candidate causal relationship, one needs to effectively determine whether the cause and effect is real or is due to the~presence of other variables in the system. A common approach is to test the relative independence between the potential cause and effect conditioned on all other variables, as demonstrated in linear models~\cite{Guo2008,Barrett2010,Papana2013}. Although analytically rigorous, this approach requires the estimation of the related inference measures for high dimensional random variables from (often limited available) data, and therefore, suffers the curse of dimensionality~\cite{Runge2012PRL}. Many heuristic alternatives have been proposed to address this issue. The essence of these alternative approaches is to repeatedly measure the relative independence between the cause and effect conditioned on a combination of the other variables, often starting with subsets of only a few variables~\cite{Marinazzo2012,Runge2012PRL,Runge2012PRE}. As soon as one such combination renders the proposed cause and effect independent, the proposed relationship is rejected, and there is no need to condition on subsets with more variables.
The advantage of such an approach is that it reduces the dimensionality of the sample space if the proposed relationship is rejected at some point. However, if the relationship is not rejected, then the algorithm will continue, potentially having to enumerate over all subsets of the other variables. In this scenario, regardless of the dimensionality of the sample space, the combinatorial search itself is often computationally infeasible for large or even moderate-size systems.

In our recent work in \cite{Sun2014PhysicaD,Sun2014arXiv}, we introduced the concept of causation entropy (CSE), proposed and proved the optimal causation entropy (oCSE) principle and presented efficient algorithms to infer causal relationships within large complex systems. In particular, we showed in \cite{Sun2014arXiv} that by a combination of the aggregate discovery and progressive removal of variables, all causal relationships can be correctly inferred in an computational feasible and data-efficient manner. We repeat for emphasis that without properly accounting for multiple interactions and conditioning accordingly, erroneous or irrelevant casual influences may be inferred, and specifically, any pairwise-only method will inherent the problem that many false-positive connections will arise. The design of oCSE specifically addresses this~issue.

In this paper, we focus on the problem of inferring the coupling structure in synthetic biological systems. When the system reaches an equilibrium state, we employ random perturbations to extract time series that approximate a linear Gaussian stochastic process. Then, we apply the oCSE principle to infer the system's coupling structure from the measured data. Finally, we show that the success rate of causal inferences not only improves with the amount of available data, but it also increases with the higher frequency of sampling.

\section{Inference of Direct Coupling Structure through Optimal Causation Entropy}\label{sec2}

To infer causal structures in complex systems, we clearly need to specify the mathematical assumptions under which the task is to be accomplished. Accurate mathematical modeling of complex systems demands taking into account the coupling of neglected degrees of freedom or, more generally, the fluctuations of external fields that describe the environment interacting with the system itself~\cite{Bazzani2003}. This requirement can be addressed in a phenomenological manner by adding noise in deterministic dynamical models. The addition of noise, in turn, generally leads to the stochastic process formalism used in the modeling of natural phenomena~\cite{Doukhan2008}. We study the system in a probabilistic framework. Suppose that the system contains $n$ components, $\mathcal{V}=\{1,2,\dots,n\}$. Each component $i$ is assumed to be influenced by a unique set of components, denoted by $N_i$. Let:
\begin{equation}
	X_t=[X^{(1)}_t,X^{(2)}_t,\dots,X^{(n)}_t]\vspace{-3pt}
\end{equation}
be a random variable that describes the state of the system at time $t$.
For a subset $K=\{k_1,k_2,\dots,k_q\}\subset\mathcal {V}$, we define:
\begin{equation}
	X^{(K)}_t\equiv[X^{(k_1)}_t,X^{(k_2)}_t,\dots,X^{(k_q)}_t].
\end{equation}
	
\subsection{Markov Conditions}

We assume that the system undergoes a stochastic process with the following Markov conditions~\cite{Sun2014arXiv}.
\begin{equation}\label{eq:processconds}
\begin{cases}
\mbox{{(i)} {Temporally Markov}:~}\\
\quad\quad\quad\quad p(X_{t}|X_{t-1},X_{t-2},\dots) = p(X_{t}|X_{t-1})= p(X_{t'}|X_{t'-1})~\mbox{for any $t$ and $t'$.}\\
\mbox{{(ii)} {Spatially Markov}:~}\\
\quad\quad\quad\quad p(X^{(i)}_{t}|X_{t-1})=p(X^{(i)}_t|X^{(N_i)}_{t-1})~\mbox{for any $i$.}\\
\mbox{{(iii)} {Faithfully Markov}:~}\\
{
\quad\quad\quad\quad p(X^{(i)}_{t}|X^{(K)}_{t-1})\neq p(X^{(i)}_t|X^{(L)}_{t-1})~\mbox{whenever
$(K\cap N_i)\neq(L\cap N_i)$.}}
\end{cases}
\end{equation}
Here, $p(\cdot|\cdot)$ denotes conditional probability. The relationship between two probability density functions $p_1$ and $p_2$ is denoted as $``p_1=p_2"$ iff they equal almost everywhere, and $``p_1\neq p_2"$ iff there is a set of positive measure on which the two functions do not equal. Note that the Markov conditions stated in Equation~\eqref{eq:processconds} ensure that for each component $i$, there is a unique set of components $N_i$ that renders the rest of the system irrelevant in making inference about $X^{(i)}$
{and each individual component in $N_i$ presents an observable cause regardless of the presence or absence of the other components.}

Although several complex systems can be properly modeled in terms of Markov processes~\cite{Eichler2012}, we cannot avoid recalling that, after all, non-Markov is the rule and Markov is only the exception in nature~\cite{Kampen1998}. Therefore, it is important to develop a theoretical framework suitable for identifying coupling structures in complex systems driven by correlated fluctuations where memory effects cannot be neglected. It is in fact possible to relax the assumptions in Equation~\eqref{eq:processconds} to a stochastic process with finite or vanishing memory, as discussed in the last section of the paper.

The problem of inferring direct (or causal) couplings can be stated as follows. Given time series data $\{x^{(i)}_t\}$ ($i=1,2,\dots,n;t=1,2,\dots,T$) drawn from a stochastic process that fulfills the Markov conditions in Equation~\eqref{eq:processconds}, the goal is to uncover the set of causal components $N_i$ for each $i$. We solve this problem by means of algorithms that implement the oCSE principle.

\subsection{Causation Entropy}

We follow Shannon's definition of entropy to quantify the uncertainty of a random variable. In particular, the entropy of a continuous random variable $X$ is defined as~\cite{Shannon1948}:
\begin{equation}\label{eq:hX}
	h(X) \equiv -\int p(x)\log p(x)dx,
\end{equation}
where $p(x)$ is the probability density function of $X$. The conditional entropy of $X$ given $Y$ is defined as~\cite{Cover}:
\begin{equation}\label{eq:hXY}
	h(X|Y)\equiv-\int p(x,y)\log p(x|y)dxdy.\vspace{3pt}
\end{equation}
Equations~\eqref{eq:hX} and~\eqref{eq:hXY} are valid for both univariate and multivariate random variables.
Consider a stochastic process.
Let $I$, $J$ and $K$ be arbitrary sets of components within the system.
We propose to quantify the influence of $J$ on $I$ conditioned upon $K$ via the CSE:\vspace{3pt}
\begin{equation}\label{eq:defcse}
	C_{J\rightarrow I|K} = \lim_{t\rightarrow\infty}[h(X^{(I)}_{t+1}|X^{(K)}_t) - h(X^{(I)}_{t+1}|X^{(K)}_t,X^{(J)}_t)],
\end{equation}
provided that the limit exists~\cite{Sun2014PhysicaD,Sun2014arXiv}.
{Since, in general, $H(X|Y)-H(X|Y,Z)=I(X;Z|Y)$, where the latter defines the conditional mutual information between $X$ and $Z$ conditioned on $Y$}, it follows from the nonnegativity of conditional mutual information that CSE is nonnegative.
When $K=\varnothing$, we omit the conditioning part and simply use the notation $C_{J\rightarrow I}$. Notice that if $K=I$, CSE reduces to TE. On~the other hand, CSE generalizes TE by allowing $K$ to be an arbitrary set of components.

\subsection{Optimal Causation Entropy Principle}

In our recent work~\cite{Sun2014arXiv}, we revealed the equivalence between the problem of identifying the causal components and the optimization of CSE. In particular, we proved that for an arbitrary component $i$, its~unique set of causal components $N_i$ is the {\it minimal set of components that maximizes CSE}.
Given the collection of sets (of components) with maximal CSE with respect to component $i$,
\begin{equation}\label{eq:optimalcse1}
	\mathcal{K}_i=\{K\subset\mathcal{V}|C_{K\rightarrow i}\geq C_{K'\rightarrow i}~\mbox{for any}~K'\subset\mathcal{V} \},\vspace{-3pt}
\end{equation}\label{eq:optimalcse2}
\noindent it was shown that $N_i$ is the unique set in $\mathcal{K}_i$ with minimal cardinality, \emph{i.e.},
\begin{equation}\label{eq:optimalcse2}
	N_i = \cap_{K\in\mathcal{K}_i}K = {\operatorname{argmin}}_{K\in\mathcal{K}_i}|K|.
\end{equation}
We refer to this minimax principle as the oCSE principle~\cite{Sun2014arXiv}.

\subsection{Computational Causality Inference}\label{sec2.4}

Based on the oCSE principle, we developed two algorithms whose joint sequential application allows the inference of causal relationships within a system~\cite{Sun2014arXiv}.
The goal of these algorithms is to effectively and efficiently identify for a given component $i$ the direct causal set of components $N_i$. The first algorithm aggregatively identifies the potential causal components of $i$. The outcome is a set of components $M_i$ that includes $N_i$ as its subset, with possibly additional components. These additional components are then progressively removed by applying the second algorithm.

For a given component $i$, the first algorithm, referred to as aggregative discovery, starts by selecting a component $k_1$ that maximizes CSE, \emph{i.e.},
\begin{equation}
	k_1={\operatorname{argmax}}_{k\in\mathcal{V}}C_{k\rightarrow i}.
\end{equation}
Then, at each step $j$ ($j=1,2,\dots$), a new component $k_{j+1}$ is identified among the rest of the components to maximize the CSE conditioned on the previously selected components:\vspace{-3pt}
\begin{equation}
k_{j+1}=\underset{k\in\mathcal{V}/\{k_1,k_2,\dots,k_j\}}{\operatorname{argmax}}C_{k\rightarrow i|\{k_1,k_2,\dots,k_j\}}.
\end{equation}
Recall that CSE is nonnegative.
The above iterative process is terminated when the corresponding maximum CSE equals zero, \emph{i.e.}, when:
\begin{equation}\label{eq:maxcsezero}
	\underset{k\in\mathcal{V}/\{k_1,k_2,\dots,k_j\}}\max C_{k\rightarrow i|\{k_1,k_2,\dots,k_j\}} = 0,
\end{equation}
and the outcome is the set of components $M_i=\{k_1,k_2,\dots,k_j\}\supset N_i$.

Next, to remove non-causal components (including indirect and spurious ones) that are in $M_i$, but not in $N_i$, we employ the second algorithm, referred to as progressive removal. A component $k_j$ in $M_i$ is removed when:
\begin{equation}\label{eq:csezero}
	C_{k_j\rightarrow i|M_i/\{k_j\}}=0,
\end{equation}
and $M_i$ is updated accordingly\footnote{An alternative way of removing non-causal components is to test conditioning on subsets of $M_i$ with increasing cardinality, potentially reducing the dimensionality of sample space at the expense of increased computational burden. This is similar to the PC-algorithm originally proposed in Ref.~\cite{Spirtes2000} and successfully employed in Ref.~\cite{Runge2012PRL}, with the key difference that here the enumeration of conditioning subsets only needs to be performed for $M_i$ instead of for the entire system $\mathcal{V}$.}.
After removing all such components, the resulting set is identified as $N_i$.

\subsection{Estimation of CSE in Practice}

In practice, causation entropy $C_{J\rightarrow I|K}$ needs to be {\it estimated} from data.
We define the Gaussian estimator of causation entropy as:
\begin{equation}\label{eq:cseest}
	\widehat{C}^{\mbox{\scriptsize (Gaussian)}}_{J\rightarrow I|K} \equiv \frac{1}{2}\log
	\left(\frac{\operatorname{det}\left[\widehat\Phi(0)_{II}-\widehat\Phi(1)_{IK}\widehat\Phi(0)_{KK}^{-1}\widehat\Phi(1)_{IK}^\top\right]}
	{\operatorname{det}\left[\widehat\Phi(0)_{II}-\widehat\Phi(1)_{I,K\cup J}\widehat\Phi(0)_{K\cup J,K\cup J}^{-1}\widehat\Phi(1)_{I,K\cup J}^\top\right]}\right).
\end{equation}
Here:
\begin{equation}
\Phi(\tau)_{IJ}\equiv\cov(X^{(I)}_{t+\tau},X^{(J)}_t)\vspace{3pt}
\end{equation}
denotes a covariance matrix where $\tau\in\{0,1\}$, and $\widehat\Phi(\tau)_{IJ}$ denotes the corresponding sample covariance matrix estimated from the data~\cite{Sun2014arXiv}. The estimation $\widehat{C}^{\mbox{\scriptsize (Gaussian)}}_{J\rightarrow I|K}\approx C_{J\rightarrow I|K}$ if: (i) the underlying random variables are Gaussians; and (ii) the amount of data is sufficient for the relevant sample covariances to be close to their true values. When the underlying random variables are non-Gaussian, an efficient estimator for CSE is yet to be fully developed. As previously pointed out, binning-related estimators, although conceptually simple, are generally inefficient, unless the sample space is low-dimensional~\cite{Schindlera2007}. To gain efficiency and to minimize bias for a large system where the sample space is high-dimensional, an~idea would be to build upon the $k$-nearest neighbor estimators of entropic measures~\cite{Vejmelka2008,Kraskov2004,Vlachos2010,Tsimpiris2012}.

Regardless of the method that is being adopted for the estimation of $C_{J\rightarrow I|K}$, the estimated value is unlikely to be exactly zero due to limited data and numerical precision. In practice, it is necessary to decide whether or not the estimated quantity should be regarded as zero. Such a decision impacts the termination criterion Equation~\eqref{eq:maxcsezero} in the aggregative discovery algorithm and determines which components need to be removed based on Equation~\eqref{eq:csezero} in the progressive removal algorithm. As discussed in \cite{Sun2014arXiv}, such a decision problem can be addressed via a nonparametric statistical test, called the permutation test, as described below.

For given time series data $\{x^{(i)}_t\}$ ($i=1,2,\dots,n;~t=1,2,\dots,T$), let $\widehat{C}_{J\rightarrow I|K}$ denote the estimated value of $C_{J\rightarrow I|K}$. Based on the set of components $J$ and a permutation function $\pi$ on the set of integers from one to $T$, the corresponding permuted time series $\{y^{(i)}_t\}$ is defined as: \vspace{3pt}
\begin{equation}
	y^{(i)}_t=
	\begin{cases}
	x^{(i)}_{\pi(t)} & \mbox{if $i\in J$,}\\
	x^{(i)}_t & \mbox{if $i\notin J$}.
	\end{cases}
\end{equation}
To apply the permutation test, we generate a number of randomly permuted time series (the number will be denoted by $r$).
We then compute the causation entropy from $J$ to $I$ conditioned on $K$ for each permuted time series to obtain $r$ values of the estimates, which are consequently used to construct an empirical cumulative distribution $F$
as:
\begin{equation}
F(C)\equiv\frac{1}{r}\big|\{\widehat{C}^{(s)}_{J\rightarrow I|K}:\widehat{C}^{(s)}_{J\rightarrow I|K}\leq C,~1\leq s\leq r\}\big|,
\end{equation}
where $\widehat{C}^{(s)}_{J\rightarrow I|K}$ are estimates from the permuted time series with $1\leq s\leq r$ and $|\cdot|$ denotes the cardinality of a set.
Finally, with the null hypothesis that $C_{J\rightarrow I|K}=0$, we regard
$C_{J\rightarrow I|K}$ as strictly positive if and only if:
\begin{equation}
	F(\widehat{C}_{J\rightarrow I|K})>\theta,\vspace{-3pt}
\end{equation}
where $0<(1-\theta)<1$ is the prescribed significance level. In other words, the null hypothesis is rejected at level $(1-\theta)$ if the above inequality holds.
Therefore, the permutation test relies on two input parameters: (i) the number of random permutations $r$; and (ii) the significance level $(1-\theta)$.
In practice, the increase of $r$ improves the accuracy of the empirical distribution at the expense of additional computational cost. A reasonable balance is often achieved when $10^{3}\lesssim r\lesssim10^4$~\cite{Sun2014arXiv}. The value of $\theta$ sets a lower bound on the false positive ratio and should be chosen to be close to one (for example, $\theta=99\%$)~\cite{Sun2014arXiv}.


\section{Extracting Stochastic Time Series from Deterministic Orbits}\label{sec3}

Biological systems can exhibit both regular and chaotic features~\cite{Murray2002}.
For instance, a healthy cardiac rhythm is characterized by a chaotic time series, whereas a pathological rhythm often exhibits regular dynamics~\cite{Garfinkel1992}. Therefore, a decrease of the chaoticity of the cardiac rhythm is an alarming clinical signature. A similar connection between the regularity of the time series and pathology is observed in spontaneous neuronal bursting in the brain~\cite{Schiff1994,Lesne2006}.
When a system settles into a periodic or equilibrium state, it becomes nearly impossible to infer the coupling structure among the variables, as the system is not generating any information to be utilized for inference. To overcome this difficulty, we propose to apply small stochastic perturbations to the system while in an equilibrium state (this is equivalent to adding dynamic noise to a system to facilitate coupling inference, as shown in \cite{Pompe2011PRE}). Then, we measure the system's response over short time intervals. Finally, we follow the oCSE principle and apply the aggregative discovery and progressive removal algorithms to the measured data to infer the couplings among the variables.

\subsection{Dynamical System and Equilibrium States}

Consider a continuous dynamical system:\vspace{-3pt}
\begin{equation}\label{eq:sys1}
	dx/dt = f(x),
\end{equation}
where $x=[x_1,x_2,\dots,x_n]^\top\in\mathbb{R}^n$ is the $n$-dimensional state variable, the symbol $``\top"$ denotes transpose, and $f=[f_1,f_2,\dots,f_n]^\top:\mathbb{R}^n\rightarrow\mathbb{R}^n$ is a smooth vector field, which models the dynamic rules of the system.
A trajectory of the system~\eqref{eq:sys1} is a solution $x(t)$ to the differential equation, Equation~\eqref{eq:sys1}, with a given initial condition $x(0)=x_0$.

An equilibrium of the system is a state $x^*$, such that $f(x^*)=0$. When a system reaches an equilibrium, the time evolution of the state ceases. An equilibrium $x^*$ is called stable if nearby trajectories approach $x^*$ forward in time, \emph{i.e.}, there exists a constant $\rho>0$, such that $x(t)\xrightarrow{t\rightarrow\infty}x^*$ whenever $\|x_0-x^*\|<\rho$, where $\|\cdot\|$ denotes the standard Euclidean norm. Otherwise, the equilibrium is called unstable.

\subsection{Response of the System to Stochastic Perturbations}

To gain information about the coupling structure of a system, it is necessary to apply external perturbations to ``knock'' the system out of an equilibrium state and observe how it responds to these perturbations.
Suppose that we apply and record a sequence of random perturbations to the system in such a manner that before each perturbation, the system is given sufficient time to evolve back to its stable equilibrium. In addition, the response of the system is measured shortly after each perturbation, but before the system reaches the equilibrium again.
Denote the stable equilibrium of interest as $x^*$; we~propose to repeatedly apply the following steps.\vspace{6pt}

Step 1. Allow the system to (spontaneously) reach $x^*$.

Step 2. At time $t$, apply and record a random perturbation $\xi$ to the system, \emph{i.e.}, $x(t)=x^*+\xi$.

Step 3. At time $t+\Delta{t}$, measure the rate of response, defined as $\eta=[x(t+\Delta{t})-x^*-\xi]/\Delta{t}$.\vspace{6pt}

Repeated application of these steps $L$ times results in a collection of perturbations, denoted as $\{\xi_\ell\}$, and rates of response, denoted as $\{\eta_\ell\}$, where $\ell=1,2,\dots,L$. Here, each perturbation $\xi_\ell$ is assumed to be drawn independently from the same multivariate Gaussian distribution with zero mean and covariance matrix $\sigma^2\mathbb{I}$.
For a given equilibrium $x^*$, in addition to $L$---the number of times the perturbations are applied---there are two more parameters in the stochastic perturbation process: the sample frequency defined as $1/\Delta{t}$ and the variance of perturbation defined as $\sigma^2$.
To ensure that the perturbation approximates the linearized dynamics of the system, we require that $1/\Delta{t}\gg 1$ and $\sigma\ll 1$. The influence of these parameters will be studied in the next section with a concrete example.

{We remark here that the choice of Gaussian distribution is a practical rather than a conceptual one. In theory, any multivariate distribution can be used to generate the perturbation vector $\xi_\ell$ as long as the component-wise distributions of $\xi^{(i)}_\ell$ and $\xi^{(j)}_k$ are identical and independent whenever $i\neq{j}$ or $\ell\neq k$. In practice, since the effectiveness of any information theoretic method (including the one proposed here) depends on the estimation of entropies, choosing the perturbations to be Gaussian greatly improves the reliability of estimation and, thus, the accuracy of the inference. This is because the entropy of Gaussian variables depends only on covariances, rendering its estimation relatively straightforward~\cite{Ahmed1989,Barnett2009PRL,Sun2014arXiv}.}

Note that each perturbation $\xi_\ell$ and its response $\eta_\ell$ are related through the underlying nonlinear differential equation $dx/dt=f(x)$, where the nonlinearity is encoded in the function $f(x)$, which is assumed to be unknown.
For an equilibrium $x^*$, the dynamics of nearby trajectories can be approximated by its linearized system, as follows. Consider a state $x\approx x^*$ and define the new variable $\delta{x}=x-x^*$. To the first order, the time evolution of $\delta x$ can be approximated by the following linear system:
\begin{equation}\label{eq:vareq}
	d(\delta{x})/dt = Df(x^*)\delta{x},
\end{equation}
where $Df(\cdot)$ is the $n$-by-$n$ Jacobian matrix of $f$, defined as $[Df(x)]_{ij}=\partial f_i/\partial x_j$. A sufficient condition for $x^*$ to be stable is that all eigenvalues of $Df(x^*)$ must have negative real parts. From Equation~\eqref{eq:vareq}, with the additional assumption that $\Delta{t}\ll 1$, the relationship between the perturbation and response is approximated by the following equation:
\begin{equation}\label{eq:linear1}
	\eta_\ell= Df(x^*)\xi_\ell.
\end{equation}
Note that since $\xi_\ell$ is a multivariate normal random variable and $Df(x^*)$ is a constant matrix, the variable $\eta_\ell$ is also (approximately) a multivariate normal random variable. Equation~\eqref{eq:linear1} therefore represents a drive-response type of Gaussian process.

\section{Application to Synthetic Biology}
\vspace{-12pt}
\subsection{The Repressilator}

Cellular dynamics is centrally important in biology~\cite{Ellner2006}. To describe and, to a certain extent, to~understand what happens in cells, fluctuating chemical concentrations and reaction rates need to be measured experimentally. However, constructing dynamic models that accurately reproduce the observed phenomena in cells is quite challenging. An alternative approach consists in engineering synthetic biological systems that follow prescribed dynamic rules. An important example is the so-called repressilator (or repression-driven oscillator) presented in \cite{Elowitz2000}.
The repressilator is based upon three transcriptional repressors inserted into the \textit{E. coli} bacteria with a plasmid. The three repressors, $lacl$, $tetR$ and $cl$, are related as follows: $lacl$ inhibits the transcription of the gene coding for $tetR$; $tetR$ inhibits the transcription of the gene coding for $cl$; $cl$ inhibits the transcription of the gene coding for $lacl$. In the absence of inhibition, each of the three proteins reaches a steady-state concentration resulting from a balance between its production and degradation rates within the bacterium. In the presence of cross-inhibition by the other two repressors, this network architecture potentially allows oscillations and other interesting dynamical behaviors and serves as a prototype of modeling the quorum sensing among bacteria species~\cite{Ullner2007}.

\begin{figure}[htbp]
\centering
\includegraphics*[width=0.8\textwidth]{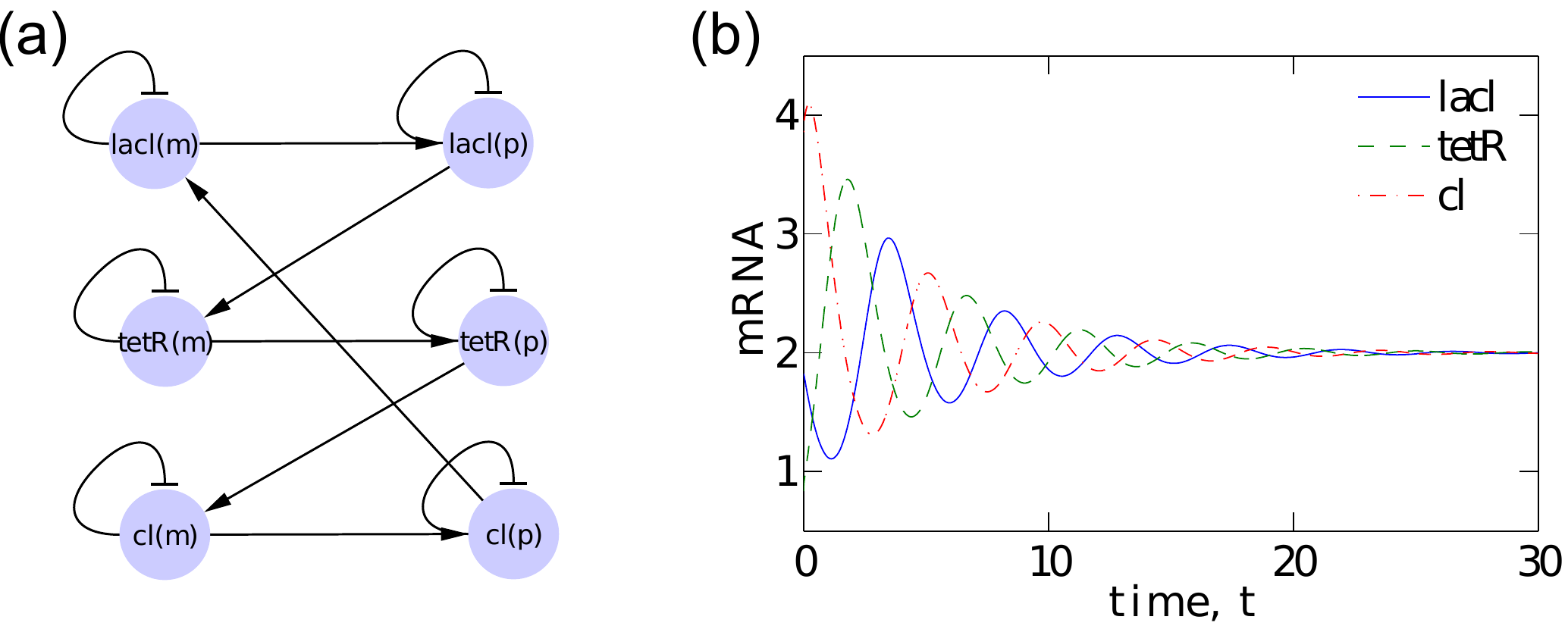}
\caption{
Illustration of the repressilator coupling structure and dynamics.
(\textbf{a}) The direct coupling structure of the repressilator as modeled by Equation~\eqref{eq:repressilator}. The system contains the mRNA and protein concentration of three genes: $lacl$, $tetR$ and $cl$. There are two types of direct couplings: positive couplings are shown as regular arrows, whereas negative (inhibitory) couplings are shown as $T$-arrows.
(\textbf{b}) Typical time series of the mRNA concentration under the parameters $n=2$, $\alpha_0=0$, $\alpha=10$ and $\beta=100$. The mRNA concentration of each gene converges to the same equilibrium state. Similar oscillatory transient and asymptotic convergence is observed for the time evolution of the protein concentration of these genes. \vspace{12pt}
}\label{fig2}
\end{figure}

The repressilator dynamics can be modeled by a system of coupled differential equations, which describe the rates of change for the concentration $p_{i}$ of each protein repressor and the concentration $m_{i}$ of its associated mRNA in their network, as:
~\cite{Elowitz2000}
\begin{eqnarray}\label{eq:repressilator}
\begin{cases}
\dot{m}_1 = -m_1 + \dfrac{\alpha}{(1+p^n_3)}+\alpha_0\\
\dot{m}_2 = -m_2 + \dfrac{\alpha}{(1+p^n_1)}+\alpha_0\\
\dot{m}_3 = -m_3 + \dfrac{\alpha}{(1+p^n_2)}+\alpha_0\\
\dot{p}_1 = -\beta(p_1-m_1)\\
\dot{p}_2 = -\beta(p_2-m_2)\\
\dot{p}_3 = -\beta(p_3-m_3)
\end{cases}
\end{eqnarray}
where $m_1$ ($p_1$), $m_2$ ($p_2$) and $m_3$ ($p_3$) represent the mRNA (protein) concentration of the genes $lacl$, $tetR$ and $cl$, respectively.
See Figure~\ref{fig2}a for a schematic representation of the system.
Each ODE in Equation~\eqref{eq:repressilator} consists of positive terms modeling the production rate and a negative term representing degradation.
There are four parameters in the ODEs in Equation~\eqref{eq:repressilator}, namely: $\beta $ is the ratio of the protein decay rate to the mRNA decay rate; $n$ is the so-called Hill coefficient and describes the cooperativity of the binding of repressor to promoter; $\alpha _{0}$, the leakiness of the promoter, is the rate of transcription of mRNA in the presence of saturating concentration of the repressor; $\alpha _{0}+\alpha $ is the additional rate of transcription of mRNA in the absence of the inhibitor. Note that units of time and concentration in Equation~\eqref{eq:repressilator} have been rescaled in order to make these equations non-dimensional~\cite{Elowitz2000}.

As shown in \cite{Elowitz2000}, there is an extended region in the parameter space for which the system described in Equation~\eqref{eq:repressilator} exhibits a single stable equilibrium. The Jacobian matrix at the equilibrium is:\vspace{6pt}
\begin{equation}
Df =
\begin{pmatrix}
-1 & 0 & 0 & 0 & 0 & -\frac{\alpha np_3^{n-1}}{(1+p^n_3)^2} \\
0 & -1 & 0 & -\frac{\alpha np_1^{n-1}}{(1+p^n_1)^2} & 0 & 0 \\
0 & 0 & -1 & 0 & -\frac{\alpha np_2^{n-1}}{(1+p^n_2)^2} & 0 \\
\beta & 0 & 0 & -\beta & 0 & 0 \\
0 & \beta & 0 & 0 & -\beta & 0 \\
0 & 0 & \beta & 0 & 0 & -\beta\vspace{-3pt}
 \end{pmatrix}.
\end{equation}\\
{\bf Problem statement:}
The goal of coupling inference is to identify the location of the nonzero entries of the Jacobian through time series generated by the system near equilibrium.

\subsection{Inference of Coupling Structure via the Repressilator Dynamics}

We consider the repressilator dynamics as modeled by Equation~\eqref{eq:repressilator} with the parameters $n=2$, $\alpha_0=0$, $\alpha=10$ and $\beta=100$. Under this setting, the system exhibits a single stable equilibrium~\cite{Elowitz2000} at which $m_1=m_2=m_3=p_1=p_2=p_3=2$. Typical time series are shown in Figure~\ref{fig2}b.
After the system settles at the equilibrium, we apply the stochastic perturbations as described in Section~\ref{sec3} and obtain a time series of the perturbations $\{\xi_\ell\}$, as well as responses $\{\eta_\ell\}$.
The goal is to identify, for each component (the mRNA or protein of a gene) in the system, the set of components that determine its dynamics (\emph{i.e.}, the variables that appear on the right-hand side of each relation in Equation~\eqref{eq:repressilator}).

To apply our algorithms according to the oCSE principle, we define the set of random variables
$X^{(i)}_t$ ($i=1,2,\dots,12; t=1,2,\dots$) as:
\begin{equation}\label{eq:maptoX}
X^{(i)}_t=
\begin{cases}
\xi^{(i)}_t & \mbox{if $1\leq i\leq 6$},\\
\eta^{(i-6)}_{t-1} & \mbox{if $7\leq i\leq 12$}.
\end{cases}
\end{equation}
The approximate relationship between the perturbation and response as in Equation~\eqref{eq:linear1} can be expressed~as: \vspace{6pt}
\begin{equation}\label{eq:linear2}
	X^{(I)}_{t+1} = AX^{(J)}_t,\vspace{3pt}
\end{equation}
where $I=\{7,8,\dots,12\}$, $J=\{1,2,\dots,6\}$ and $A=Df(x^*)$.
Since this equation defines a stochastic process that satisfies the Markov assumptions in Equation~\eqref{eq:processconds}, the oCSE principle applies. This implies that the direct couplings can be correctly inferred, at least in principle, by jointly applying the aggregative discovery and progressive removal algorithms.

Since the perturbations $\{\xi_\ell\}$ and responses $\{\eta_\ell\}$ depend on the number of samples $L$, the rate of perturbation $1/\Delta{t}$ and the variance of perturbation $\sigma^2$, so is the accuracy of the inference.
Next, we explore the change in performance of our approach by varying these parameters.
We use two quantities to measure the accuracy of the inferred coupling structure, namely, the false positive ratio $\varepsilon_{+}$ and the false negative ratio $\varepsilon_{-}$.
{Since our goal is to infer the structure rather than the weights of the couplings, we focus on the structure of the Jacobian matrix $Df(x^*)$, encoded in the binary matrix $B$, where:
\begin{equation}
[B]_{ij}=
\begin{cases}
1, & \mbox{if $[Df(x^*)]_{ij}\neq 0$},\\
0, & \mbox{otherwise}.
\end{cases}
\end{equation}
On the other hand, applying the oCSE principle, the inferred direct coupling structure gives rise to the estimated binary matrix $\widehat{B}$, where:
\begin{equation}
[\widehat{B}]_{ij}=
\begin{cases}
1, & \mbox{if $j$ is a direct causal component of $i$},\\
0, & \mbox{otherwise}.\vspace{-6pt}
\end{cases}
\end{equation}
Given matrices $B$ and $\widehat{B}$, the false positive and false negative ratios are defined, respectively, as:
\begin{equation}\label{eq:inferror}
\begin{cases}
\varepsilon_{+}\equiv \dfrac{\mbox{number of $(i,j)$ pairs with $\widehat{B}_{ij}=1$ and $B_{ij}=0$}}{\mbox{number of $(i,j)$ pairs with $B_{ij}=0$}},\vspace{0.1in}\\
\varepsilon_{-}\equiv \dfrac{\mbox{number of $(i,j)$ pairs with $\widehat{B}_{ij}=0$ and $B_{ij}=1$}}
{\mbox{number of $(i,j)$ pairs with $B_{ij}=1$}}.
\end{cases}
\end{equation}
It follows that $0\leq\varepsilon_{+},\varepsilon_{-}\leq1$ and $\varepsilon_{+}=\varepsilon_{-}=0$ when exact (error-free) inference is achieved.}

\begin{figure}[htbp]
\centering
\includegraphics*[width=0.8\textwidth]{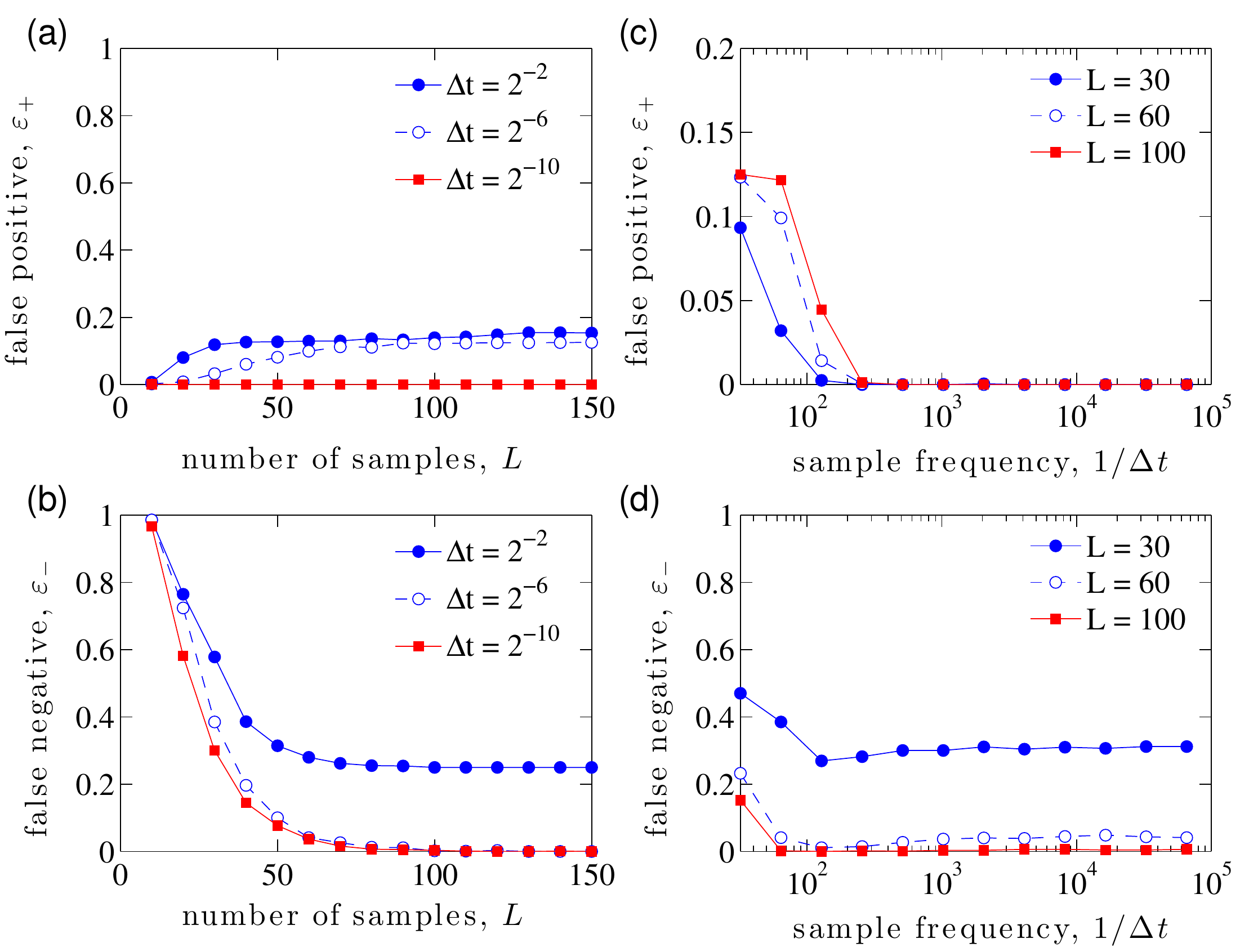}
\caption{
Direct inference of couplings in the repressilator system. We consider the repressilator dynamics modeled by Equation~\eqref{eq:repressilator} with the parameters $n=2$, $\alpha_0=0$, $\alpha=10$ and $\beta=100$.
The system poses a single stable equilibrium.
Given $L$ (number of samples), $\sigma^2$ (variation of perturbation) and $\Delta{t}$ (lag time), we apply stochastic perturbations, as described in Section~\ref{sec3}, and obtain time series of the perturbations $\{\xi_\ell\}$ and responses $\{\eta_\ell\}$. The time series are then converted into $\{x_t\}$ according to Equation~\eqref{eq:maptoX}, and direct couplings are inferred via the aggregative discovery and progressive removal algorithms (see Section~\ref{sec2.4} for details).
(\textbf{a},\textbf{b}) False positive $\varepsilon_{+}$ and false negative $\varepsilon_{-}$ as a function of $L$ for three different values of $\Delta{t}$. Here, $\varepsilon_{+}$ and $\varepsilon_{-}$ are defined in Equation~\eqref{eq:inferror}.
(\textbf{c},\textbf{d}) $\varepsilon_{+}$ and $\varepsilon_{-}$ as a function of $1/\Delta{t}$ for three different values of $L$.
In all panels, we set $\sigma=10^{-2}$, and each data point is an average over $100$ independent runs.  \vspace{12pt}
}\label{fig3}
\end{figure}

\begin{figure}[htbp]
\centering
\includegraphics*[width=0.8\textwidth]{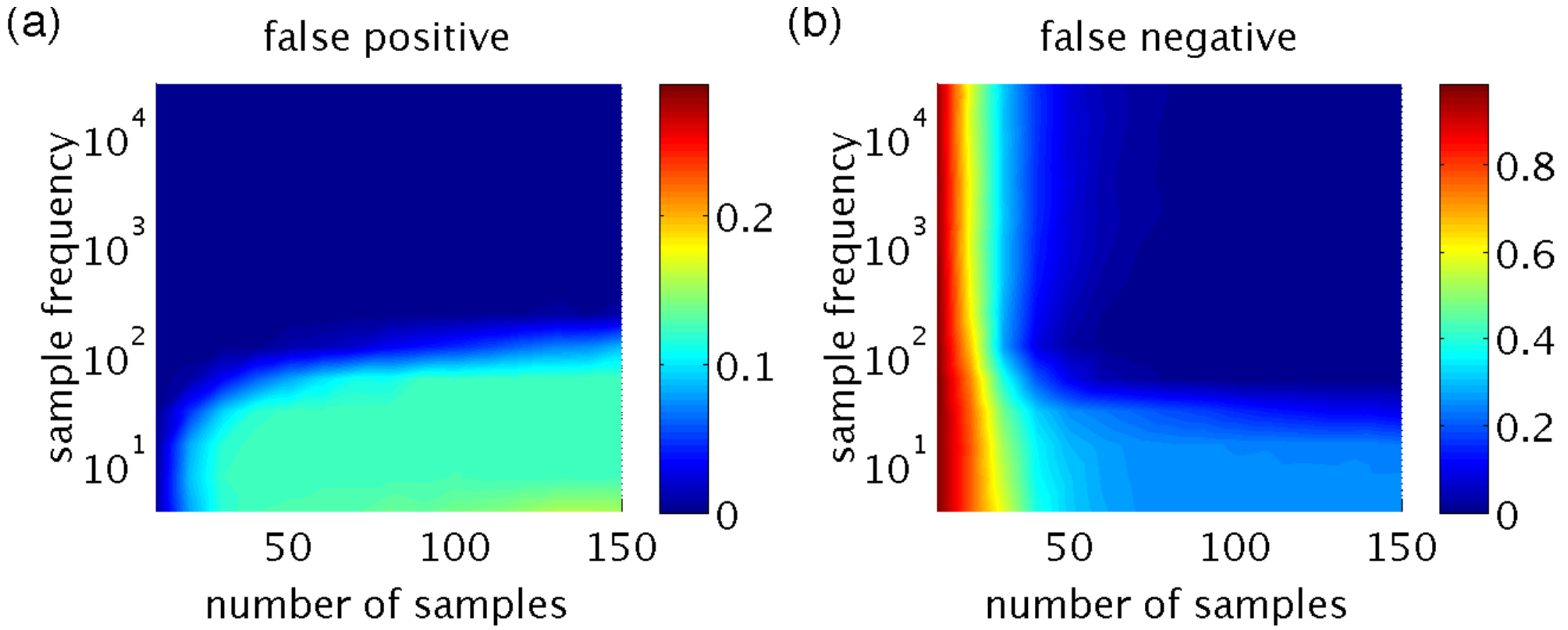}
\caption{
The same setup as in Figure~\ref{fig3} (for fixed $\sigma=10^{-2}$), shown as surface plots.\vspace{12pt}
}\label{fig4}
\end{figure}

Figure~\ref{fig3}a,b shows that both the false positives and false negatives converge as the number of samples increases. However, they converge to zero only if the rate of perturbation is sufficiently high. Figure~\ref{fig3}c,d supports this observation and, in addition, shows that in the high rate of perturbation regime, exact inference is achieved with a sufficient number of samples (in this case, $L\sim 100$). The combined effects of $L$ and $1/\Delta{t}$ are shown in Figure~\ref{fig4}.
In all simulations, the variation of the perturbation is set to be $\sigma^2=10^{-4}$ and is found to have little effect on the resulting inference, provided that it is sufficiently small to keep the linearization in Equations~\eqref{eq:linear1} and \eqref{eq:linear2} valid.

\section{Discussion and Conclusions}
\vspace{-12pt}
\subsection{Results Summary}

In this paper, we considered the challenging problem of inferring the causal structure of complex systems from limited available data (enjoy Figure~\ref{fig1} for a cartoon depiction of the concept of causality during a soccer game). Specifically, we presented an application of the so-called oCSE principle to identify the coupling structure of a synthetic biological system, the repressilator. First, we briefly reviewed the main points of the oCSE principle (Equations~\eqref{eq:optimalcse1} and \eqref{eq:optimalcse2}), which states that for an arbitrary component $i$ of a complex system, its unique set of causal components $N_{i}$ is the minimal set of components that maximizes CSE (causation entropy, defined in Equation~\eqref{eq:defcse} is a generalization of transfer entropy). We strengthen in this work our claim that CSE is a suitable information-theoretic measure for reliable statistical inferences, since it takes into account both direct and indirect influences that appear in the form of information flows between nodes of networks underlying complex systems. We also devoted some attention to the implementation of the oCSE principle. This task is accomplished by means of the joint sequential application of two algorithms, aggregative discovery and progressive removal, respectively. Second, having introduced the main theoretical and computational frameworks, we used a stochastic perturbation approach to extract time series data approximating a Gaussian process when the model system---the repressilator (see Equation~\eqref{eq:repressilator})---reaches an equilibrium configuration. We then applied the above-mentioned algorithms implementing the oCSE principle in order to infer the coupling structure of the model system from the observed data. Finally, we numerically showed that the success rate of our causal entropic inferences not only improves with the amount of available measured data (Figure~\ref{fig2}a), but it also increases with a higher frequency of sampling (Figure~\ref{fig2}b).

One especially important feature of our oCSE-based causality inference approach is that it is immune (in principle) to false positives. When data is sufficient, false positives are eliminated by sequential joint application of the aggregative discovery and progressive removal algorithms, as well as raising the threshold $\theta$ used in the permutation test (see \cite{Sun2014arXiv} for a more detailed investigation of this point). In contrast, any pairwise causality measure without additional appropriate conditioning will in principle be susceptible to false positives, result in too many connections and, sometimes, even, implies that everything causes everything. Such a limitation is intrinsic and cannot be overcome merely by gathering more data~\cite{Sun2014arXiv}. On the other hand, false negatives are less common in practice and are usually caused by ineffective statistical estimation or\linebreak insufficient data.

\subsection{oCSE for General Stochastic Processes}

We presented the oCSE principle for stochastic processes under the Markov assumptions stated in Equation~\eqref{eq:processconds}.
From an inference point of view, each and every one of these assumptions has a well-defined interpretation. For instance, the temporal Markov condition implies the stationarity of causal dependences between nodes. The loss of temporal stationarity could be addressed by partitioning the time series data into stationary segments and, then, performing time-dependent inferences. Clearly, such inferences would require extra care. Furthermore, the loss of the spatially and/or faithfully Markov condition would imply that the set of nodes that directly influence any given node of the network describing the complex system is no longer minimal and unique. Causality inference in this case becomes an ill-posed problem. These issues will be formally addressed in a forthcoming work.

Note that a finite $k$-th order Markov process $\{X_t\}$ can always be converted into a first-order (memoryless) one $\{Z_t\}$ by lifting, or in other words, defining new random variables \linebreak $Z_t=(X_{t-k+1},X_{t-k+2},\dots,X_t)$~\cite{Gallager2013}.
In this regard, the oCSE principle extends naturally to an arbitrary finite-order Markov process. On the other hand, for a general stationary stochastic process that is not necessarily Markov (\emph{i.e.}, considering a process with potentially infinite memory), there might exist an infinite number of components from the past that causally influence the current state of a given component. However, under an assumption of vanishing (or fading) memory, such influences decay rapidly as a function of the time lag, and consequently, the process itself can be approximated by a finite-order Markov chain~\cite{Harris1955,Berbee1987,Fernandez2002,Fernandez2005}.
We plan to leave such generalizations for forthcoming investigations.


{\center\bf Acknowledgments}
We thank Samuel Stanton from the Army Research Office (ARO) Complex Dynamics and Systems Program for his ongoing and continuous support.
This work was funded by ARO Grant No. W911NF-12-1-0276.

{\center\bf Author Contributions}
Jie Sun and Erik M. Bollt designed and supervised the research.
Jie Sun and Carlo Cafaro performed the analytical calculations and the numerical simulations.
All authors contributed to the writing of the paper.

{\center\bf Conflicts of Interest}
The authors declare no conflict of interest.



\end{document}